\documentclass[preprint,showpacs,aps]{revtex4}
\usepackage{amsmath,amssymb,epsfig}
\begin{document} 
\title[Soliton-like base pair opening  in a helicoidal  model of DNA ]
{Soliton-like base pair opening  in a helicoidal   DNA: An analogy with helimagnet and cholesterics }

\author{M. Daniel }
\email{daniel@cnld.bdu.ac.in}
\author{
V. Vasumathi}
\email{ vasu@cnld.bdu.ac.in}
\affiliation{ Centre for Nonlinear Dynamics, School of Physics, 
Bharathidasan University, Tiruchirappalli - 620 024, India.
}

\date{\today} 
\begin{abstract}
We propose a  model  for DNA dynamics by introducing the helical structure  through twist deformation in analogy with the structure of  helimagnet and cholesteric liquid crystal system.
The  dynamics in this case is found to be governed by the completely integrable sine-Gordon equation which admits kink-antikink solitons with increased width representing a wide base pair opening configuration in DNA. The results show that the helicity introduces a length scale variation and thus  provides a better representation of the base pair opening in DNA.           
\end{abstract}

\maketitle
 
  The B-form  DNA double helix molecule 
  is  usually modeled by two parallel chains of nucleotides known as strands  
   with  linkage interms of 
  dipole-dipole  interaction   along the strands and the two strands 
  are coupled to each other through 
  hydrogen bonds between the  complementary bases \cite{ref10}. Molecular excitations in DNA based on the above model is generally governed by nonlinear evolution
  equations \cite{ref1,ref2,ref3} and in particular by the completely integrable sine-Gordon-type equations
 \cite{ref6,ref8}.   In the above studies, DNA is treated  as  two coupled linear chains 
  without involving the helical character of its structure.
  However, in nature DNA exists in a double helix form and recently there were
   attempts by few authors to study the dynamics by taking into account the
   helical character  of the double helix through  different forms of coupling. For instance, Gaeta \cite{ref14,refg3,refg4}, 
  Dauxios \cite{ref16} and Cadoni et al \cite{ref15a}  assumed that  the torsional  coupling between the
  $n^{th}$  base on one strand and the $(n+4)^{th}$ base on the complementary strand is the  responsible force for
  the helical nature in DNA and  found that the  localized excitations are governed by solitons and breathers.  Barbi et al \cite{ref17a,ref17b} and Campa \cite{ref18} however introduced the helicity through a proper choice of
  the coupling between the radial and the angular variables of the helix and  obtained    breathers and kinks.  On the other hand, very recently, Takeno \cite{ref19} introduced  helicity in DNA through a helical transformation and  obtained  non-breathing compacton-like modes to represent base pair opening through numerical calculations.\\
 In this, paper, we propose a  model by  introducing  the helical character in each strand  of the DNA molecule   through a twist deformation  of the chain in analogy with the twist in  cholesteric  liquid crystal \cite{ref20} or  orientation of spins in a helimagnet \cite{ref21}. 
\begin{figure}
\begin{center}
\epsfig{file=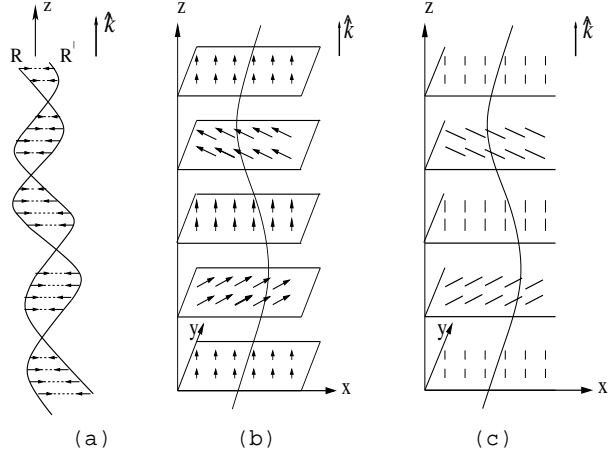,height =6cm, width=8cm}
\caption{A schematic representation  of (a) a DNA double helical chain (b) a helimagnet  and (c) a cholesteric liquid crystal system }
\end{center}
\end{figure}
 As an illustration in  Fig. 1(a-c) we have presented a schematic representation of the arrangement of bases, spins and molecules respectively in a DNA double helical chain, in a helimagnet  and in a cholesteric liquid crystal leading to the formation of helical structure. In Fig. 1(a) $R$ and $R' $  represent the two complementary strands of the DNA double helix and the dots between the arrows represent the  hydrogen bonds  between the complementary bases. The arrows and short lines   in Figs. 1(b) and 1(c)  respectively  represent the spins  and molecules  at different sites  and planes in a helimagnet and in a cholesteric liquid crystal. When we go along the z-direction, the orientation of spins  and molecules are tilted from one plane to the next through certain tilt  angle.  If we join the tips of the arrows representing the  spin vectors and also, the tips of the molecules they form a helix as shown in Figs. 1(b) and 1(c) respectively.\\

 In a recent paper,
one of the present authors   studied the nonlinear spin dynamics of a helimagnet by incorporating the helicity interms of  Frank free energy corresponding to the twist deformation which is responsible for helicity in a cholesteric liquid crystal system \cite{ref21,refmd}. The  Frank free energy density associated with the twist deformation in a cholesteric liquid crystal is given by $[{\bf p\cdot(\nabla\times  p)}-q_0]^2$ where the unit vector ${\bf p}$ represents the director axis which corresponds to the average direction of orientation of the liquid crystal molecules, $q_0=\frac{2\pi}{q}$ is the pitch wave vector and  $q$  is the pitch of the helix.  The discretised form of the above twist free energy is written as $\{[\hat k\cdot{\bf( p_n\times  p_{n+1})}]-q_0\}^2$ where $\hat k$ is the unit vector along z-direction. In analogy with the above, we write down  the free energy associated with the twist deformation  in terms of spin vector   as $\{[\hat k\cdot{\bf(S_n\times S_{n+1})}]-q_0\}^2$. By taking into account the form of free energy the Heisenberg model of  Hamiltonian for an anisotropic helimagnetic system is written as \cite{ref21}   
\begin{eqnarray}
 H_1=\sum_n\left[-J({\bf  S_n \cdot  S_{n+1}})+A ({S_n}^{z})^{2}\right.\nonumber\\
\left.+h\{[\hat k\cdot{\bf (S_n\times S_{n+1})}]-q_0\}^2\right].\label{eq1}
\end{eqnarray}
In Eq.(\ref{eq1}), ${\bf S_n}=(S_n^x,S_n^y,S_n^z)$ represents the spin vector at the $n^{th}$ site and the terms proportional to $J$ and $A$ respectively represent the ferromagnetic spin-spin exchange interaction and uniaxial magneto-crystalline anisotropy with the easy axis along z-direction.
 $h$ denotes the elastic constant associated with the twist deformation. We identify the above helical spin chain with one of the strands of the DNA double helical chain. Therefore, in a similar fashion we can write down the spin Hamiltonian $H_2$  for another helimagnetic system corresponding to the complementary strand with the spin vector ${\bf S_n}$ replaced by ${\bf S'_n}$. We  assume that in the Hamiltonian the exchange, anisotropic and twist coefficients as well as the pitch  in  both the helimagnetic systems are equal.  Now, for mapping the helimagnetic spin system with the DNA double helical chain we   rewrite the Hamiltonian by writing  the spin vectors as   ${\bf
S_n}\equiv(S_n^{x}, S_n^{y}, S_n^{z})=(\sin\theta_n\cos\phi_n,   \sin\theta_n\sin\phi_n,  \cos\theta_n)$ and  $ {\bf S'_n}\equiv(S_n^{'x}, S_n^{'y}, S_n^{'z})=(\sin\theta'_n\cos\phi'_n,  \sin\theta'_n\sin\phi'_n,  \cos\theta'_n)$, where $\theta_n (\theta'_n)$ and $\phi_n(\phi'_n)$ are the angles of rotation of spins in the xy and xz-planes respectively. The new Hamiltonian corresponding to $H_1$ is written as 

\begin{eqnarray}
H_1&=&\sum_n\left[-J \{ \sin\theta_n\sin\theta_{n+1} \cos
(\phi_{n+1}-\phi_n)\right.\nonumber\\
&&+ \cos\theta_n\cos\theta_{n+1}\}+A\cos\theta_n^{2}+h\{\sin\theta_n\nonumber\\
&&\left.\times\sin\theta_{n+1}\sin(\phi_{n+1}-\phi_n)-q_0\}^2\right],\label{eq4}
\end{eqnarray}

We now map the two helical spin systems with the two strands of the DNA double helix with the two angles $\theta_n (\theta'_n)$ and $\phi_n(\phi'_n)$ representing the angles of rotation of bases in the xz and xy-planes of the two strands respectively. A horizontal projection of the $n^{th}$ base of  DNA in the xy and xz-planes is shown in Figs. 2(a) and 2(b).  Here $Q_n$ and $Q'_n$ denote the tips of the $n^{th}$ bases attached to the strands $R$ and $R'$ at $P_n$ and $P'_n$ respectively. The DNA double helix chain is stabilised by stacking of bases through intrastrand dipole-dipole interaction and through hydrogen bonds (interstrand interaction) between complementary bases. The interstrand 
     base-base interaction or hydrogen bonding energy between the complementary bases  
      depends on  the distance between them and using the simple geometry in Figs. 2(a,b), we can write the distance between the tips of bases as \cite{ref8},
\begin{eqnarray}
 ( Q_nQ'_n)^{2}&\approx &2\left[\sin\theta_n\sin\theta'_n
\left(\cos\phi_n\cos\phi'_n+\sin\phi_n\sin\phi'_n\right)\right.\nonumber\\
&&\left.-\cos\theta_n\cos\theta'_n\right].
\end{eqnarray} 
\begin{figure}      
\begin{center}
\epsfig{file=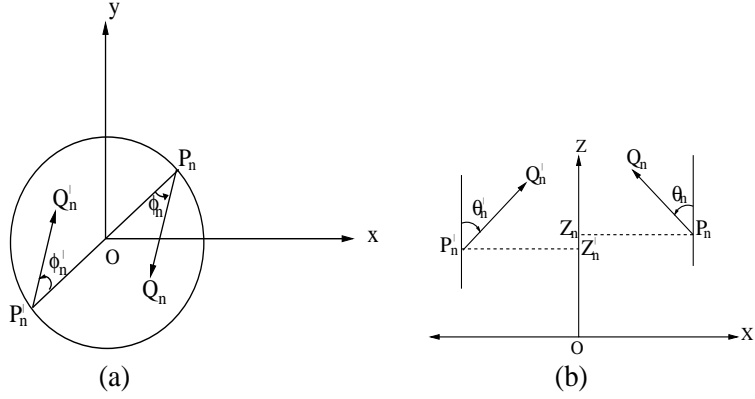,height =6cm, width=10cm}
\caption{ A horizontal projection of the $n^{th}$ base
pair in a DNA double helix (a) in the xy-plane and (b)  in the xz-plane.}
\end{center}
\end{figure}  
Now,  the above equation represents the hydrogen bonding energy between complementary bases and we can write down the Hamiltonian for the interstrand interaction or hydrogen bonds  is written as
\begin{eqnarray}
 H_{12}&=&\eta\left[\sin\theta_n\sin\theta'_n
\left(\cos\phi_n\cos\phi'_n+\sin\phi_n\sin\phi'_n\right)\right.\nonumber\\
&&\left.
-\cos\theta_n\cos\theta'_n\right],\label{eq6a}
\end{eqnarray}
where $\eta$ is a constant.
The total Hamiltonian  for our helicoidal model of DNA interms of the angles of rotation of bases  using the above Hamiltonians is written  as
\begin{eqnarray}
H&=& H_1+H_2+H_{12}\nonumber\\
&=&\sum_n\left[-J \{ \sin\theta_n\sin\theta_{n+1} \cos
(\phi_{n+1}-\phi_n)+ \cos\theta_n\right.\nonumber\\
&&\times\cos\theta_{n+1}+\sin\theta'_n
\sin\theta'_{n+1}\cos(\phi'_{n+1}-\phi'_n)\nonumber\\
&& + \cos\theta'_n\cos\theta'_{n+1}\}+h\{[sin\theta_n\sin\theta_{n+1}
\nonumber\\
&&\times\sin(\phi_{n+1}-\phi_n)-q_0]^2+[\sin\theta'_n\sin\theta'_{n+1}\nonumber\\
&&\times\sin(\phi'_{n+1}-\phi'_n)-q_0]^2\}+\eta\{\sin\theta_n\sin\theta'_n
\nonumber\\
&&\times\left(\cos\phi_n\cos\phi'_n+\sin\phi_n\sin\phi'_n\right)-\cos\theta_n\cos\theta'_n\}\nonumber\\
&&\left.+A(\cos^{2}\theta_n+{\cos^{2}\theta'_n})\right].
\label{eq6b}
\end{eqnarray}
 Using the equation of motion for the corresponding quasi-spin model \cite{refjt} in the limit $A>>J,\eta,h$,  we obtain  $ \dot \phi_n= 2 A \cos\theta_n$ and $ \dot\phi'_n= 2 A'\cos\theta'_n $. Hence, under absolute minima of
 potential  the  Hamiltonian (\ref{eq6b}) becomes 
\begin{eqnarray}
 H&=&\sum_n\left[ \frac{I}{2}( {\dot\phi_n}^{2}+ {\dot\phi_n}^{'2}) +J 
 [2-\cos (\phi_{n+1}-\phi_n)\right.
 \nonumber\\
&&-\cos  (\phi'_{n+1}-\phi'_n)]-\eta [1
  -\cos  (\phi_n-\phi'_n)]\nonumber\\
&&+h\{2q_0^2-[\sin
 (\phi_{n+1}-\phi_n)-q_0]^2\nonumber\\
&&\left.-[\sin
 (\phi'_{n+1}-\phi'_n)-q_0]^2\}\right], \label{eq7}
 \end{eqnarray}
where $I=\frac{1}{2A^2}$ is the moment of inertia of  the bases around the axes at $p_n (p'_n)$.
While rewriting the   Hamiltonian in the above form, we have restricted that the bases are rotating in the plane which is normal to the helical axis. In otherwords, we have now restricted our problem to a plane base rotator model \cite{ref8} by assuming  $\theta=\theta'=\pi/2$.\\ 

Having formed the Hamiltonian, the dynamics of the DNA double helix molecule can be understood by constructing the Hamilton's equations of motion corresponding to the Hamiltonian (\ref{eq7})  as 
  \begin{subequations}
  \begin{eqnarray}
   I\ddot\phi_n&=& \left[J+2h\cos(\phi_{n+1}-\phi_n)\right] \sin (\phi_{n+1}-\phi_n)\nonumber\\
&&-\left[J+2h\cos(\phi_n-\phi_{n-1})\right]\sin
  (\phi_n-\phi_{n-1}) \nonumber\\
&&  
+\eta  \sin (\phi_n-\phi'_n)-2hq_0[\cos(\phi_{n+1}-\phi_n)\nonumber\\
&&-\cos(\phi_n-\phi_{n-1})], \label{eq9a}\\
  I\ddot\phi'_n&=& \left[J+2h\cos(\phi'_{n+1}-\phi'_n)\right] \sin (\phi'_{n+1}-\phi'_n)\nonumber\\
&&-\left[J+2h\cos(\phi'_n-\phi'_{n-1})\right]\sin
  (\phi'_n-\phi'_{n-1})\nonumber\\
&&  
+\eta  \sin (\phi'_n-\phi_n)-2hq_0[\cos(\phi'_{n+1}-\phi'_n)\nonumber\\
&&-\cos(\phi'_n-\phi'_{n-1})], \label{eq9b}  
  \end{eqnarray}  
  \end{subequations} 
where overdot represents derivative with respective to time.
 Eqs. (\ref{eq9a}) and (\ref{eq9b}) describe the dynamics of the DNA double helix  at the discrete level when the helical nature of the molecule is represented in the form  of a twist-like deformation. \\

 It is expected that the difference in angular rotation of bases with respect to neighbouring bases along the two strands is small \cite{ref6,refsm}.  Also, very recently Gaeta \cite{refg3,refg4} proposed that the helical structure of DNA will introduce qualitative changes only in the small amplitude regime. Hence, under the small angle approximation,  in the continuum limit,  the discrete equations of motion (7a,b) after suitable rescaling of time and redefinition of the parameter $\eta$ reduce to

\begin{subequations}
\begin{eqnarray}
\phi_{ t t}&=& \frac{(J+2h)}{I}\phi_{zz}+ \eta \sin(\phi-\phi'),\label{eq11a}\\
\phi'_{ t t}&=&\frac{(J+2h)}{I} \phi'_{zz} + \eta\sin(\phi'-\phi).\label{eq11b}
\end{eqnarray}
\end{subequations}
 Adding and subtracting
 Eqs. (\ref{eq11a}) and (\ref{eq11b})  and after suitable rescaling of the variable $z$,   we obtain
\begin{eqnarray}
\Psi_{ t t}-\Psi_{ z z}+\sin\Psi=0,\label{eq13}
\end{eqnarray}
where $\Psi=2\phi$ and we have further chosen $2\eta=-1$.  Also, while  deriving  Eq. (\ref{eq13}), we have chosen  $\phi'=-\phi$, because among the possible rotations of bases, rotation of complementary bases in opposite direction easily facilitate an open state configuration.  Eq. (\ref{eq13}) is the completely integrable sine-Gordon equation which was originally solved for N-soliton
      solutions interms of kink and antikink using the most celebrated Inverse Scattering Transform  method  \cite{ref23}.  For instance, the kink and antikink one  soliton solution of the sine-Gordon equation
 is written interms of the original variables as
\begin{eqnarray}
\phi(z,t)&=&2\arctan[exp[\pm \frac{1}{\sqrt{1-v^2}}\nonumber\\
&&\times \sqrt{\frac{I}{(J+2h)}} (z -v  t)]],\label{eq14}
\end{eqnarray}
where $+$ and $-$ represent the kink and antikink soliton solutions respectively and $v$ is the velocity of the soliton. In Fig. 3(a) we have plotted  the angular  rotation of bases $\phi$   in terms of the kink-antikink one soliton solution as given in Eq. (\ref{eq14}) by choosing the stacking, helicity, moment of inertia and velocity parameters respectively as $J=1.5~ eV,~ h=3.0~ eV,~ I =1.3\times 10^{-36} g~cm^2 $ and $v=0.4 ~cm ~s^{-1}$ \cite{ref16,ref19}.
\begin{figure}
\begin{center}
\epsfig{file=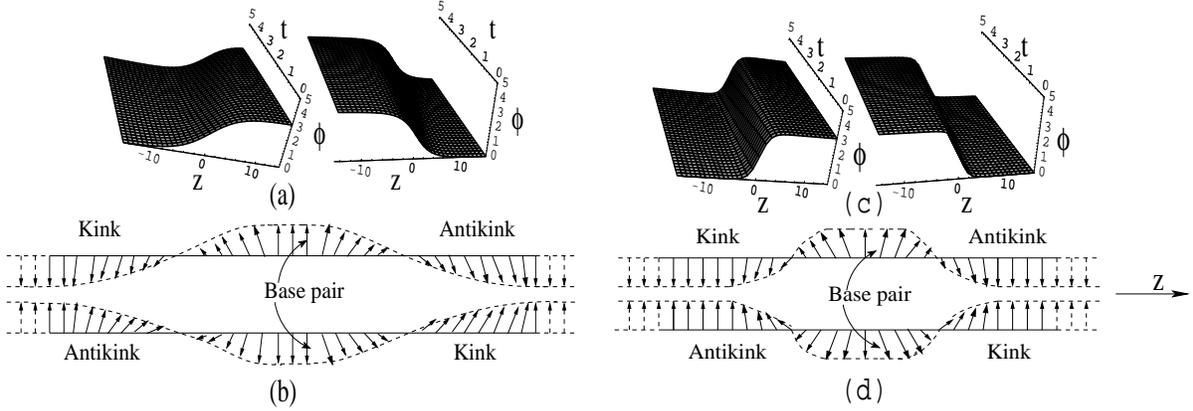,height =6cm, width=16cm}
\caption{(a) Kink  and antikink one soliton solutions (Eq. (\ref{eq14}))  of the sine-Gordon equation  when helicity is present
$(h\neq 0)$.
 (b) A sketch of the formation of open state configuration in terms of
kink-antikink solitons in a DNA double helix when helicity  is present $(h\neq 0)$.
(c) Kink  and  antikink one soliton solutions  of the sine-Gordon equation  when helicity is absent
(Eq. (\ref{eq14}) when $  h= 0$).
 (d) A sketch of the formation of open state configuration in terms of
kink-antikink solitons in DNA double helix when helicity is absent $(h= 0)$.}
\end{center}
\end{figure} 
\begin{figure}
\begin{center}
\epsfig{file=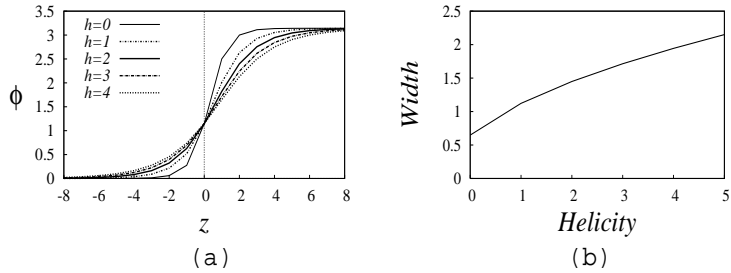,height =3.8cm, width=10cm}
\caption{(a)  The kink   one soliton  (Eq. (\ref{eq14})) representing base pair opening  at   $t=1$ for different  values of helicity. 
 (b)  Variation of  width of the  kink soliton against helicity.
 }
\end{center}
\end{figure} 
The kink-antikink soliton solution which can propagate infinite distance and time describes an open
       state  configuration in the DNA double helix  which is schematically represented in Fig. 3(b).  In order to understand the effect of helicity on the open state configuration, in Fig. 3(c), we  have also plotted the kink and antikink one soliton solution of the sine-Gordon equation in the absence of  helicity, that is by choosing $h=0$ (keeping all other parameters values the same), in the solution given in Eq.(\ref{eq14}). From Figs. 3(a) and 3(c), we observe that when there is  helicity in the model ($h\neq 0$), the kink-antikink soliton is getting broadened.  In other words,  helicity in DNA makes more number of base pairs to participate in the formation of open state configuration without introducing any qualitative change in the dynamics. This is also schematically represented in Fig. 3(d) which looks evident on comparing  Fig. 3(b). In order to highlight the above fact,   we have  separately plotted the kink one soliton solution at a given time ($t=1$) for different  values of helicity by choosing $h=0,1,2,3,4$  in Fig. 4(a).  The  increase in width against  helicity is explicitly represented in Fig. 4(b).  From the figure it may be noted that the increase in helicity slows down the rotation of bases and makes more and more  number of base pairs to participate in the open state configuration, thus providing a better representation of base pair opening in DNA. Thus, helicity introduces a length scale variation in the base pair opening.\\
  Similar results have also been observed  by Dauxios \cite{ref16} through a perturbation analysis  on his helicoidal model of DNA and obtained soliton with a  much broader width. In order to have a more realistic model,  dissipative (viscous effect) and noise terms should be added to the equations of motion. Experimentally, the life time  of soliton in this case is shown to be of few nano seconds at room temperature (see for e.g.  Ref.\cite{mp}).  Also, in a recent paper, Yakushevich et al \cite{yaku} through a numerical analysis,  showed that when the viscosity is strong the soliton moves a length of only few chain links and it will stop after that. On the other hand, when  the viscosity is low the soliton  passes more than 3000 chain links like  a heavy Brownian particle which is  found to be  stable with respect to thermal oscillation. When the above two effects are taken into account Eq. (\ref{eq13})  takes the form $\Psi_{ t t}-\Psi_{ z z}+\sin\Psi=\epsilon [\alpha \Psi_t+\beta\digamma (z,t)]$ where the terms proportional to $\alpha$ and $\beta$ are related to viscous surrounding and thermal forces respectively. A soliton perturbation  analysis \cite{mdvv} on the above equation shows  that when the viscosity is high the soliton moves for a small distance and then stops. But when the viscosity is low the soliton moves for a long distance along the chain. The detailed analytical calculations of the above study will be separately published elsewhere.

In summary, we proposed a new  helicoidal model to study DNA dynamics by introducing the helical character  in analogy with the twist deformation  in a cholesteric  liquid crystal system and  the spin arrangement in a helimagnet. The nonlinear dynamics of  DNA  under the present  helicoidal model is found to be governed by  the completely integrable sine-Gordon equation in the continuum limit which admits kink and antikink soliton solutions. From the nature of solitons, we observe that   helicity introduces a length scale variation without causing any change in the shape of the soliton. Due to this scaling variation,  the width of the soliton increases and hence we obtain broader kink-antikinks.  In otherwords a large number of bases  are involved  in  the base pair opening, thus leading to a better representation.  This broadened base pair opening  may act as a better energetic activator in the case of RNA-Polymerase transport during transcription process in DNA. As the continuum helicoidal model does not introduce qualitative changes in the DNA dynamics, we propose to study the full nonlinear dynamics of the helicoidal model of DNA (without making small angle approximation) by solving equations (\ref{eq9a}) and (\ref{eq9b}) numerically and the results will be published elsewhere. \\ 
The work of M. D  and V.V   form part of a major
  DST  project.     


\begin{thebibliography}{99}
\bibitem{ref10}  
  L. V. Yakushevich, Nonlinear Physics of DNA (Wiley-VCH, Berlin, 2004).
\bibitem{ref1}
S. W. Englander, N. R. Kallenbanch, A. J. Heeger, J. A. Krumhansl and S. Litwin,    Proc. Natl. Acad. Sci. U.S.A {\bf 77}, 7222 (1980). 
\bibitem{ref2}
M. Peyrard and A. R. Bishop,  Phys. Rev. Lett. {\bf 62},  2755 (1989). 
\bibitem{ref3}
V. Muto, P.S. Lomdahl and P.L. Christiansen,  Phys. Rev. A {\bf 42}, 7452 (1990).
\bibitem{ref6}
S. Yomosa, Phys. Rev. A  {\bf 27}, 2120 (1983); {\bf 30}, 474 (1984).
\bibitem{ref8}
S. Takeno and S. Homma, Prog. Theor. Phys. {\bf  70}, 308 (1983); {\bf 72}, 679 (1984).
\bibitem{ref14}
G. Gaeta, Phys. Lett. A {\bf 143}, 227 (1990); {\bf 168}, 383 (1992).
\bibitem{refg3}
G. Gaeta,  Phys. Rev .E {\bf 74}, 021921 (2006).
\bibitem{refg4}
G. Gaeta, J. Nonlin. Math. Phys. {\bf 14}, 57 (2007).
\bibitem{ref16}
T. Dauxios,  Phys. Lett. A {\bf 159}, 390 (1991).
\bibitem{ref15a}
M. Cadoni, R. De Leo and G. Gaeta, Phys. Rev. E {\bf 75}, 021919 (2007).
\bibitem{ref17a}
M. Barbi, S. Cocco and M. Peyrard,  Phys. Lett. A   {\bf 253}, 358 (1999).
\bibitem{ref17b}
M. Barbi, S. Lepri , M. Peyrard and N. Theodorakopoulos,  Phys. Rev. E   {\bf 68}, 061909 (2003).
\bibitem{ref18}
 A. Campa,  Phys. Rev. E  {\bf 63}, 021901 (2001).
\bibitem{ref19}
S. Takeno,  Phys. Lett. A  {\bf 339}, 352 (2005).
\bibitem{ref20}
P. G. de Gennes, {\it Physics of Liquid Crystals} (Clarendon Press, Oxford) (1974).
\bibitem{ref21}
R. M. White,  {\it Quantum Theory of Magnetism} (Springer, New York) (1982).
\bibitem{refmd}
M. Daniel and J. Beula,    {\it Choas Soliton and Fractals} (Accepted for Publication) (2008).
\bibitem{refjt}
J. Tjon and J. Wright, Phys. Rev. B {\bf 15}, 3470 (1977).
\bibitem{refsm}
G. Fa Zhou,   Physica Scripta {\bf 40}, 694 (1989).
\bibitem{ref23}
M. J. Ablowitz, D. J. Kaup, A. C. Newell, and H. Segur, Stud. Appl. Math.  {\bf 53}, 249 (1974).
\bibitem{mp}
M. Peyrard, Nonlinearity, {\bf 17}, R1 (2004) and references therein.
\bibitem{yaku}
L. V. Yakushevich, A. V. Savin and L. I. Manevitch, Phys. Rev. E {\bf 66},  016614 (2002).
\bibitem{mdvv}
M. Daniel and V. Vasumathi, Physica D {\bf 231}, 10 (2007).
\end{thebibliography}
 \end{document}